\def\year{2021}\relax
\documentclass[letterpaper]{article} 
\usepackage{aaai21}  
\usepackage{times}  
\usepackage{helvet} 
\usepackage{courier}  
\usepackage[hyphens]{url}  
\usepackage{graphicx} 

\def\UrlFont{\rm}  
\usepackage{natbib}  
\usepackage{caption} 
\begin{document}

\title{Preclinical Stage Alzheimer's Disease Detection Using Magnetic Resonance Image Scans}

\author {

        Fatih Altay \textsuperscript{\rm 1} \thanks{Equal Contribution},
        Guillermo Ramón Sánchez \textsuperscript{\rm 1} \footnotemark[1],
        Yanli James \textsuperscript{\rm 2},
        Stephen V. Faraone \textsuperscript{\rm 2},
        Senem Velipasalar \textsuperscript{\rm 1},
        Asif Salekin \textsuperscript{\rm 1} \thanks{Corresponding author}\\
}
\affiliations {
    \textsuperscript{\rm 1} Syracuse University \\
    \textsuperscript{\rm 2} SUNY Upstate Medical University \\
    faltay@syr.edu, grsanche@syr.edu, zhangy@upstate.edu, sfaraone@childpsychresearch.org, svelipas@syr.edu, asalekin@syr.edu

}
\maketitle

\begin{abstract}
Alzheimer's disease is one of the diseases that mostly affects older people without being a part of aging. The most common symptoms include problems with communicating and abstract thinking, as well as disorientation. It is important to detect Alzheimer's disease in early stages so that cognitive functioning would be improved by medication and training. In this paper, we propose two attention model networks for detecting Alzheimer's disease from MRI images to help early detection efforts at the preclinical stage. We also compare the performance of these two attention network models with a baseline model. Recently available OASIS-3 Longitudinal Neuroimaging, Clinical, and Cognitive Dataset is used to train, evaluate and compare our models. The novelty of this research resides in the fact that we aim to detect Alzheimer's disease when all the parameters, physical assessments, and clinical data state that the patient is healthy and showing no symptoms.
\end{abstract}

\section{Introduction}
Alzheimer's disease (AD) is a type of brain disease, which is degenerative, and its symptoms worsen over the years. According to the Alzheimer’s Association \footnotemark, 5.8 million Americans are living with AD. Statistics have also shown that annual AD related death rates by age are increasing every year \cite{mortrate}. According to \cite{20years}, AD could begin twenty years or more before symptoms are perceptible, and between one and six years with changes in the brain that are unnoticeable to the person affected. As the disease evolves, more and more brain neurons stop functioning, lose connection, even die. At first, AD affects the entorhinal cortex and hippocampus brain regions that are involved in memory \cite{zott2018happens}. Eventually, it affects the cerebral cortex, which is responsible for language, reasoning, and social behavior, and many other areas of the brain.
\par 

This pathology, according to the Alzheimer's Association \footnotemark[\value{footnote}], is the most common cause of dementia, especially among older people. AD accounts for 60-80\% of the total dementia cases \cite{dementia}. AD significantly reduces the quality of life \cite{logsdon1999quality} and life expectancy \cite{rice2001prevalence} of patients, and is considered as the most expensive disease in USA \cite{hurd2013monetary}.
When the economic impact of this disease is considered, just in 2019, the total combined payments from all AD patients are estimated at \$290 billion \footnotemark[\value{footnote}].

\footnotetext{https://www.alz.org/media/documents/alzheimers-facts-and-figures-2019-r.pdf}

\par
While there is no cure for Alzheimer’s disease or a way to stop its progression after diagnosis, and the treatment of AD is still an open research question, there are drug and non-drug options that may help treat symptoms. Studies have demonstrated that early stage intervention of AD can significantly impact the degeneration process, and treatment of symptoms \cite{earlytratfinds,golde2018alzheimer,crous2017alzheimer}. The early detection of AD through conventional MRI scanning will facilitate effective and in-time interventions/treatments, that would expand the life expectancy and quality of life of patients. 
\par

National Institute on Aging has defined three stages of AD: (1) Preclinical, when the patients do not exhibit any symptoms, but brain neuronal structure has started to deteriorate; (2) Mild cognitive impairment (MCI), when patients start to exhibit cognitive impairments, but still can perform all activities of daily living (ADL); and (3) Alzheimer’s dementia, when symptoms of dementia are severe enough to interfere with ADLs.
\par 
With the advancement of MRI technology \cite{lohrke201625}, and the recent development of deep learning based computer vision approaches, several studies have addressed the detection of AD in MCI and dementia stages (stage 2 and 3) from MRI brain scans \cite{10.1117/12.2281808,Bohle}. While these studies have improved our understanding of Alzheimer’s disease, they do not contribute in early stage interventions, since stage 2 and 3 are accurately identifiable through clinical diagnosis.
Accurate detection or indication of preclinical AD is a major interest in the medical community \cite{schindler2017neuropsychological,perneczky2018biomarkers}. However, to the best of our knowledge, existing studies have not addressed the challenge of preclinical AD detection from MRI brain scans yet.


Recent reports on AD suggest that \cite{sperling2011testing} change in brain may be evident 20 years before the stage of dementia (stage 3), and that there is already substantial neuronal loss by the stage of mild cognitive impairment (MCI) (stage 2). Hence, the goal of this study is the development of an effective machine learning approach that can identify the latent patterns due to preclinical AD from MRI brain scans, which can significantly improve intervention and treatment of AD patients.



MRI brain scans are 3 dimensional (3D) image representation of the brain structure. Several sequential classifiers such as, 3D Convolutional Neural Network (CNN) and 3D recurrent visual attention (RVN) model \cite{Neuro-Dram}, have been implemented to detect disease related patterns from the 3D brain scans. Recently, Transformer \cite{transformer} model has been demonstrated to outperform all the existing sequential classifiers. 
\par 
In this study, we employ and adapt two attention network models (3D recurrent visual attention model \cite{Neuro-Dram} and Transformer model \cite{transformer}) to our problem of preclinical AD detection from 3D MRI brain scan data, and compare their performances with a baseline model that is based on 3D CNNs. We evaluated these approaches for differentiating individuals with`preclinical AD' from `others'. The `others' class includes healthy individuals or patients suffering from other dementia problems not related with AD. According to our evaluation, the 3D CNN model achieves an F1 score of $0.83$ and accuracy of 88.34\%, the 3D RVN model achieves an F1 score of $0.87$, and 90.65\% accuracy, and the developed transformer model achieves an F1 score of $0.90$ and 91.18\% accuracy in preclinical  (i.e., prior to MCI and dementia stage) Alzheimer disease detection. Experimental results demonstrate that, by using the MRI images effectively, it is possible to detect preclinical stage Alzheimer's disease with a very promising accuracy. 

\par 

\section{Related Work}
Deep learning algorithms perform very well in identifying complex structures in high dimensional data, and this is why there is a very rich literature on detecting diseases using magnetic resonance images (MRI) and positron emission tomography (PET). \cite{7950647} studied the presence of Alzheimer's disease using two different 3D CNN approaches, namely VoxCNN and residual neural networks (RNN), on MRI data. Also aiming for binary classification, Alzheimer's/Non Alzheimer's, \cite{10.1117/12.2281808} used two 3D CNNs on MRI and PET scans, combining them with a fully connected layer and a softmax classification. 

Over the past few years, there has been a trend in applying attention based approaches to medical problems. With these kind of models, we can not only detect a certain pathology, but also represent which part of the data is more important to make accurate predictions. \cite{SCHLEMPER2019197} proposed a novel method, referred to as attention gated networks, for medical image analysis. This algorithm learns to focus on target structures, and can be used for leveraging certain regions for classification purposes. With a similar goal, \cite{Neuro-Dram} proposed Neuro-Dram, a 3D recurrent visual attention, for explainable neuroimaging classification. \cite{Bohle} used a layer-wise relevance propagation (LRP) to visualize CNN decisions based on MRI data. This algorithm attributes relevance to every input node, and studies the contribution of each node. It can output a heat map to highlight the most informative parts of every input image.   

Early detection of AD can be critical when developing an optimal treatment for each patient \cite{earlytreat}. \cite{earlytratfinds} showed how  early stage AD patients, improved their general cognitive abilities after 12 weeks of paper-based cognitive training. Along these lines, there has been a lot of work trying to differentiate between mild cognitive impairment (MCI) and Alzheimer's disease (AD). According to \cite{Jour}, MCI causes a slight and measurable decline in cognitive abilities and it is the earliest clinically detectable stage before AD. Suffering from MCI increases the risk of developing AD. \cite{10.3389/fncom.2019.00072} proposed a four-class SVM classifier: AD, MCI stable (patients with MCI who do not develop AD), MCI converted (patients whose MCI develop into AD), and healthy patients. Similarly, to study MCI to AD conversion, \cite{earlystage} proposed a multi-modal recurrent neural network. In this case, they used MRI, demographic information, cerebrospinal fluid (CSF) biomarkers and cognitive performance reports as their inputs to the GRU units. With a slightly different approach, \cite{enfin} used a model inspired by Inception-V4 network \cite{incep} to detect between non demented, very mild dementia, mild dementia and moderate dementia.

The goal of all the discussed approaches is detecting AD (stage 3) or predicting MCI cases (stage 2) that will develop AD, i.e. existing studies focus on stages 2 and 3. To the best of our knowledge, there is no previous work on detecting preclinical stage AD (stage 1) when all the indicators, including the clinical assessments by doctors, ensure that the patient is healthy and no symptoms of the disease are present. Our main goal is to detect future AD from the latent brain scan patterns even before MCI develops, when the disease is in a preclinical stage.

\section{Dataset}
In this work, we employ the recently published longitudinal neuroimaging, clinical and cognitive dataset, called OASIS-3 \cite{oasis3}. It consists of MRI and PET imaging from 1098 individuals collected across several studies over the course of 15 years. There are 605 cognitively normal adults and 493 individuals at different stages of cognitive decline. Ages of the participants range from 42 to 95 years. The dataset contains over 2000 MRI sessions. 

For every patient, the aging and disability resource center (ADRC) clinical data is available. This data includes patient's age at entry, height, weight, and clinical dementia rating (CDR). Diagnoses for this datatype include ``cognitively normal", ``AD dementia", ``vascular dementia" and factors that could be contributing such as vitamin deficiency, alcoholism, and mood disorders. 
The goal of this study is to identify potential AD patients even when they were clinically diagnosed as ``cognitively normal'' from the latent patterns of their brain scans. Hence, in this work, we are only using brain imaging and the affiliated clinical diagnoses, matching each scan with the closest clinical diagnosis available.
\par

The presented classifiers perform a binary classification task: differentiating preclinical AD individuals from others. More specifically, our presented binary classifiers use `0' for the patients diagnosed as healthy or with other non-Alzheimer's related pathologies, and `1' for the patients with `preclinical stage AD'. We define a `preclinical stage AD' patient as a person who is currently diagnosed by a doctor as healthy but we know that in the future he or she will develop AD. For this, we label each MRI session independently, matching every clinical data that we have with the closest MRI session available. 
\par
In this dataset, MRI brain scans are available from two different views, the axial or horizontal plane,  and sagittal or longitudinal plane \cite{planes}. We decided to use the axial plane since only a smaller portion of the individual's data contain sagittal scans. Since not all the images had the same resolution, as part of our image pre-processing, we resized and normalized all the images. 
\par 
One of the initial challenge we encountered was the significant class imbalance. More specifically, for class `0' (healthy subjects), there were 2181 scans, whereas for class `1' (patients with preclinical stage AD), we had only 176 scans. To address this issue, we down-sampled our class `0' and over-sampled the class `1', by randomly rotating and mirroring some of the images (as proposed in \cite{oversampling} and \cite{downsampling}). We also implemented a balanced sampling process. This method creates a sampler in the data loader based on the number of images of each class that yields the next index/key set to fetch. This is useful to ensure that we have an even number of labels on each training batch. 

\section{Proposed Approach}
\begin{figure*}[t!]
\centering
\includegraphics[width=0.8\textwidth]{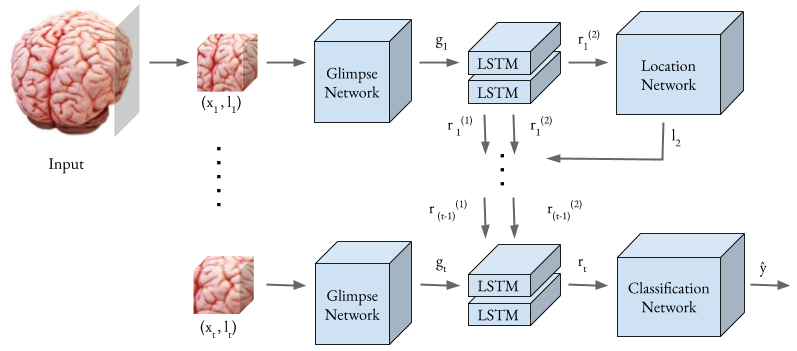} 
\caption{Overview of 3D Recurrent Visual Attention Model}
\label{fig1}
\end{figure*}
As mentioned above, existing works addressed stage 2 and 3 AD detection. One of the significant differences of our work, compared to the state-of-the-art, is that we address the more challenging problem of preclinical AD (stage 1) detection. To ensure the latent brain pattern extraction for preclinical AD detection task, we only considered the 3D brain scans as input. This makes the classification task more challenging, since we cannot adapt the models to different factors that could be critical when predicting AD, such as the age or the sex of the patient. To address this challenging problem, we adapt and employ two different attention mechanisms, namely 3D recurrent visual attention model, and attention transformer. We compare our two attention mechanisms with a baseline model, which is based on 3D CNNs.  

\subsection{Baseline - 3D CNN Model}

The baseline model is based on a 3D CNN model, which was initially used for video classification tasks \cite{hhtseng}. This model uses 3D kernels and channels to convolve video input, where the videos are viewed as 3D data (2D images over time dimension). For our baseline model, we stack all the images in a brain scan, turn them into 3D input data, and then feed it to the network. The model we developed consists of five convolutional layers and three fully connected layers. Each convolution layer is followed by batch normalization, ReLu, dropout, and pooling layers. We consider this model as our baseline.

\subsection{3D Recurrent Visual Attention Model}
For our first model shown in Fig.~\ref{fig1}, we employ a recently proposed 3D recurrent visual attention model, which is tailored for neuroimaging classification \cite{Neuro-Dram} and focuses on already developed AD detection task. This model uses a recurrent attention mechanism \cite{glimpse} that tries to find relevant locations of brain scan indicative of AD. 
The model consists of an agent that is trained with reinforcement learning. It is built around a two-layer recurrent neural network (RNN). At each timestamp, the agent receives a small portion of the entire image, which is a glimpse, centered around a position $l$, and decides which location to select at the next timestamp. After a fixed number of steps, a classification decision is made. The aim of using an agent is to maximize the rewards along the timestamps, and then decide to attend the most informative regions of the images. We define our reward,  $r_t$, as $1$ for all timestamp $t$ if the classification is correct, or $0$ if it is not. Overall, the model consists of four different networks: the glimpse network, the recurrent network, the location network and the classification network.

\textbf{The glimpse network} takes a small 3D image fraction $x_t$, and its location coordinate $l_t$ as input, and outputs a vector $g_t$. It generates a representation of the glimpse (i.e., 3D image fraction) summarizing the `what' $g_{x_t}$, and `where' $g_{l_t}$ information. 
The glimpse network consists of 3D convolutional layers (with batch normalization and max pooling) that generates the `what' representation, and a single-layer fully connected layer that converts the location coordinated to the `where' representation.
Final $g_t$ is obtained by an element-wise multiplication of these representations: 
\begin{equation} \label{eq1}
    g_t  = {g_{x_t}} \odot  {g_{l_t}}
\end{equation}

\textbf{The recurrent network} is used to obtain the agent's internal representation encapsulating the information extracted from past timestamps. This network consists of two stacked LSTM units \cite{LSTM}.
At each timestamps, hidden-layer representation generated by the LSTM units is fed into the location network to obtain the next timestamp's glimpse location, $l_{t+1}$. At the last timestamp (i.e., last iteration of the LSTM sequential analysis) the hidden layer representation is fed into the classification network.


\textbf{The location network} consists of a single-layer fully connected layer, which maps $r_{(t)}^2$ to a 3D vector in the range [-1,1] which is an isotropic 3D normal distribution. Next location $l_{t+1}$ is then produced by sampling from this distribution.

Finally, \textbf{the classification network} consists of a single fully connected layer with a sigmoid activation function, which is used for binary classification. As previously stated, the input to this network is the final timestamp hidden layer representation of the LSTM unit.
\par 
Total number of timestamps/iteration $T$ in the recurrent network, and the glimpse size are hyper-parameters. In our evaluation we identified $T=6$, and the glimpse size (3D image fraction) $40\times 40\times 40$ as the beneficial value, based on the grid search on possible values.

\subsection{Attention Transformer}

\begin{figure*}[t]
\centering
\includegraphics[width=1\textwidth]{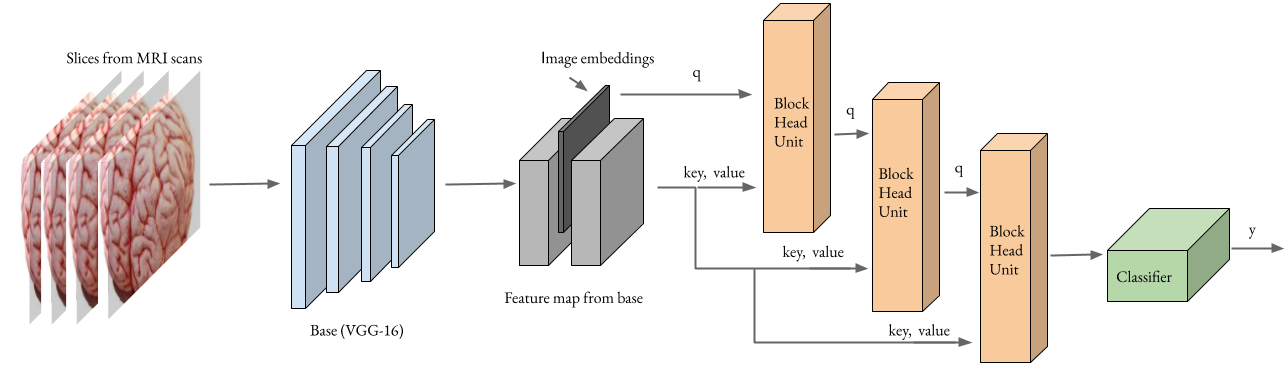} 
\caption{Overview of Attention Transformer Model}
\label{fig3}
\end{figure*}

As our second model, we employed a transformer network for the task of preclinical AD detection. Transformer models have been used for different tasks such as human action recognition from videos \cite{transformer} and text translation \cite{originaltransformer} . Although transformer networks have been used for other tasks and applications, we firmly believe that this is the first work that employs a transformer network on MRI images of brain for preclinical stage Alzheimeir's disease detection. Slices from a brain scan are fed to the network, and the network is expected to detect whether any sign of dementia is observable or not, even the subject is showing no signs nor symptoms of the disease yet.

Implemented Transformer model consists of a base network and a head network similar to \cite{transformer}. Base network extracts feature-representations from each of the brain MRI image slices, and the head network generates the binary inferences (i.e., preclinical stage AD or others).
Both the base and the head networks are described along with the modifications we have made in the following subsections. 

\subsubsection{Base Network.}

Since we deal with binary classification for preclinical stage AD detection, we use a VGG16 network \cite{vgg16} as our base network. We extract frames from brain MRI scans (96 frames) 
and feed them into the base network. Due to the nature of the base network, it accepts $224\times 224 \times 3$ images, where $224\times 224$ represents image height and weight, and 3 represents the channel size. On the other hand, slices from brain MRI scans differ in size, and are gray scale, which has only one channel. After resizing our input images to $224\times 224$, the first modification we do for the base network is adding an extra convolutional layer at the beginning to get the desired input for the base network. We set in\_channels=1, out\_channels=3, kernel\_size=3, stride=1, padding=1 and dilation=1 for this new convolutional layer. 

We have experimented with two different approaches, namely training the base network from scratch and using a pre-trained model (trained on conventional RGB images). As detailed in the experiments section, we obtained better results when the base network was trained from scratch. Instead of stacking images, (compared to 3D CNN and 3D recurrent visual attention model), we feed each of the brain scan images/frames in a sequence, like a video. The base network portion of the Fig. \ref{fig3} illustrates the network described so far. We then send the output of the base network to the head network.

\subsubsection{Transformer Unit.} 

The original transformer architecture was proposed in \cite{originaltransformer} for sequence to sequence tasks to overperform recurrents models. It is done by selecting a feature frame and comparing it with all features in a sequence in order to compute attention. To do that, features are mapped to a query (Q) and memory (K for key and V for value) embeddings using linear projections. Since the original transformer is designed for language to language translation, Q is the word that is being translated and K and V are the linear projections of the generated input and output sequences.

On the other hand, our transformer network consists of a positional encoder, three head units,  shown as Block Head Units in Fig.~\ref{fig3}, and a classifier. This unit takes the brain MRI image sequence features and also the positional embeddings for attention proposal regions and maps them into query (Q) and memory features (K, V). Query (Q) represents the region with signs of preclinical stage AD. Frames around that region are projected into K and V. Block Head Units as in Fig. \ref{fig3} process the query and memory embeddings to update the query vector. By doing so, they aggregate the information over the brain MRI scans to classify whether the given sequence belongs to preclinical stage AD class or not. 

We evaluated using different optimizers, and SGD \cite{SGD} was the most beneficial choice. We set learning rate as $1e^{-4}$, momentum as $9e^{-4}$ and nesterov as True \cite{pmlr-v28-sutskever13}. We use cross-entropy loss \cite{cross-entropy} for our classifier.

\section{Experiments}
We first split our dataset into three person-disjoint subsets: training, validation and test subsets. When trying to detect a certain pathology using patients' scans, it is crucial to split the data correctly. Different scans from the same person cannot appear among different datasets. If so, there is a risk that the model gets trained to correctly label those repeated patients, and it will perform poorly on scans from never-before-seen patients. This is why we decided to use person disjoint datasets. Each participant's MRI scan consists of 256 images. 
After the downsampling and oversampling process previously described, we used 65\% of the available data for training, 20\% for validation and 15\% for testing.



\begin{figure}[h!]
\centering
\includegraphics[width=0.4\textwidth]{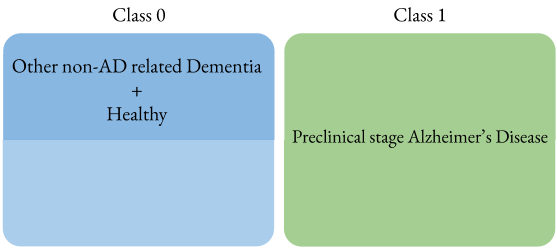} 
\caption{Overview of our data labeling approach}
\label{fig:venndiagram1}
\end{figure}

It should be emphasized that the goal of this study is to detect scans, which were confirmed to be healthy by the doctors at the time of the scan, and which we know that in the future will develop  Alzheimer's disease. Hence, we developed binary classifiers with preclinical stage Alzheimer's patients labeled as class `1', and the healthy patients or patients suffering from other dementia problems not related with Alzheimer's labeled as `0'. The evaluations discussed in this section follow the data labeling scheme shown in Fig.~\ref{fig:venndiagram1}.
In line with the study goal, we exclude the scans of patients, who have already developed and been diagnosed with AD, from the `other' category (i.e., class $0$), in order to focus on better differentiating between healthy and preclinical AD patients. As previously stated, to the best of our knowledge, there is no previous work on detecting preclinical stage AD.

For each model, we calculate the confusion matrix, F1 score, false negative rate, precision, recall, and accuracy. One of the best metrics to compare models when dealing with an imbalanced classification problem is the F1 score \footnote{https://towardsdatascience.com/metrics-for-imbalanced-classification-41c71549bbb5}. In the following subsections, we will present the obtained results for each model, and compare them in terms of the F1 score. Also, since our main task is being able to detect possible AD patients when they are still healthy, the false negative rate will also be considered as an important comparative metric.

\subsubsection{Results for the 3D CNN model.}
We used our 3D CNN model as our baseline. The best version of this model is trained with a learning rate of $1e^{-5}$ for $50$ epochs. To avoid overfitting the training data, we chose AdamW \cite{AdamW} as our optimizer with a weight decay of $0.1$. We used cross-entropy~\cite{cross-entropy} as our loss function. With these configurations, we obtained the confusion matrix shown in Table \ref{tab:CM_baseline}.

\begin{table}[h]
\resizebox{\columnwidth}{!}{%

\begin{tabular}{cc|c|c|}
\cline{3-4}
                                                 &                                                                    & \multicolumn{2}{c|}{Actual}                                                                                                   \\ \cline{2-4} 
\multicolumn{1}{c|}{}                            & Categories                                                         & \begin{tabular}[c]{@{}c@{}}Healthy /\\ other dementia\end{tabular} & \begin{tabular}[c]{@{}c@{}}Preclinical\\ AD\end{tabular} \\ \hline
\multicolumn{1}{|c|}{Predicted} & \begin{tabular}[c]{@{}c@{}}Healthy /\\ other dementia\end{tabular} & 99                                                                 & 15                                                       \\ \cline{2-4} 
\multicolumn{1}{|c|}{}                           & \begin{tabular}[c]{@{}c@{}}Preclinical\\ AD\end{tabular}           & 4                                                                  &   45                                                       \\ \hline
\end{tabular}
}
\caption{Confusion matrix for the baseline model.}
\label{tab:CM_baseline}
\end{table}

According to the evaluation, the 3D CNN model achieves an F1 score of $0.83$ on preclinical stage AD detection. 
As shown in Table \ref{tab:CM_baseline}, we have a false negative (FN) rate of $25$\%. By taking a closer look at some of the patients wrongly labeled as healthy or other dementia when they were preclinical AD patients, we identified that they are the patients whose clinical data is not very consistent. More specifically, there are some patients who are diagnosed with AD at some point of the data collection process, and then they recover their cognitive abilities. Therefore, these patients are diagnosed as ``Cognitive normal" for the rest of their life. We find that this type of patients are the hardest to correctly classify. Additionally, OASIS 3 dataset contains brain scans from patients with non-Alzheimer (non-AD) related dementia, that 3D CNN baseline model finds difficult to classify correctly. On the other hand, the baseline model performs well in classifying subjects who, once diagnosed with AD dementia, do not fully recover their cognitive abilities. 

As an example, the clinical data of patient ID 30205 (on OASIS 3 dataset) is shown in Table \ref{tab:patient30205}. The patient was clinically diagnosed as ``Cognitively normal’’ on day 0000, 0406, and 0773. OASIS 3 Dataset contains a complete MRI brain scan of the patient performed at day 61. Our baseline 3D CNN model can  detect `preclinical stage AD’ from these scans performed on day 61. Thus, the baseline algorithm is able to detect `preclinical stage AD' 1064 days (1125 - 61  = 1064) before the patient is  diagnosed with uncertain dementia. If we just consider when the patient is actually diagnosed with Alzheimer's disease, we are able to detect that pathology 1,776 days (1837-61) before it is diagnosed. 

\begin{table}[h]
\centering
\begin{tabular}[h]{lcc}
    \hline
     Day & Diagnose \\
    \hline
    0000 & Cognitively normal  \\
    0406  & Cognitively normal \\
    0773  & Cognitively normal \\
    1125  & uncertain dementia  \\
    1460  & uncertain dementia \\
    1837  & Alzheimer's disease dementia \\

    \hline
\end{tabular}
\caption{Clinical data for patient 30205.}
\label{tab:patient30205}
\end{table}

\subsubsection{Results for the 3D recurrent visual attention model.}

With the attention mechanism models incorporated, our goal is to outperform the baseline. For the 3D recurrent visual attention (3D RVN) model, the best performance is achieved with a learning rate of $1e^{-4}$, trained for $200$ epochs using AdamW \cite{AdamW}. We also evaluated different glimpse sizes (i.e., image fraction size). More specifically, we experimented with $20\times 20 \times 20$, $40\times 40 \times 40$, and $60\times 60\times 60$. The best results presented in Table \ref{tab:CMglimpse} were obtained with the $40\times 40 \times 40$ glimpse size. According to the evaluation, the 3D RVN model achieves an F1 score of $0.87$ for `preclinical stage AD' detection, which is 5\% higher compared to the baseline 3D CNN model. Additionally, according to the results, the false negative (FN) rate is $19.2$\%, which is lower than the 3D CNN model.

\begin{table}[h]

\resizebox{\columnwidth}{!}{%
\begin{tabular}{cc|c|c|}
\cline{3-4}
                                                 &                                                                    & \multicolumn{2}{c|}{Actual}                                                                                                   \\ \cline{2-4} 
\multicolumn{1}{c|}{}                            & Categories                                                         & \begin{tabular}[c]{@{}c@{}}Healthy /\\ other dementia\end{tabular} & \begin{tabular}[c]{@{}c@{}}Preclinical\\ AD\end{tabular} \\ \hline
\multicolumn{1}{|c|}{Predicted} & \begin{tabular}[c]{@{}c@{}}Healthy /\\ other dementia\end{tabular} & 84                                                                 & 10                                                       \\ \cline{2-4} 
\multicolumn{1}{|c|}{}                           & \begin{tabular}[c]{@{}c@{}}Preclinical\\ AD\end{tabular}           & 3                                                                  & 42                                                       \\ \hline
\end{tabular}
}
\caption{Confusion matrix for the 3D RVN model.}
\label{tab:CMglimpse}
\end{table}

\par 
As an example, the clinical data of patient ID 30025 (on OASIS 3 dataset) is shown in Table \ref{tab:patient30025}. The patient was clinically diagnosed as ``Cognitively normal’’ from day 0000 to day 2608, and diagnosed as ``AD dementia’’ on day 2933. OASIS 3 Dataset contains two complete MRI brain scans of the patient performed on day 0210 and 2298. Our 3D RVN model correctly infers `preclinical stage AD’ by taking the brain scan taken on day 0210 as input. 
This result demonstrates that
the presented 3D RVN can detect AD 2723 days before it is clinically diagnosed by doctors. 

\begin{table}[h]
\centering
\begin{tabular}[h]{lcc}
    \hline
     Day & Diagnose \\
    \hline
    0000 & Cognitively normal  \\
    0359  & Cognitively normal \\
    0751  & Cognitively normal \\
    1106  & Cognitively normal  \\
    1547  & Cognitively normal\\
    1915  & Cognitively normal\\
    2247  & Cognitively normal\\
    2608  & Cognitively normal\\
    2933  & Alzheimer's disease dementia \\

    \hline
\end{tabular}
\caption{Clinical data for patient 30025.}
\label{tab:patient30025}
\end{table}

One of the important benefits of this model is that not only we can detect AD before any symptoms, but also we can plot the part of the data that the algorithm interprets as more important when making accurate decisions. 
In our implementation, the initial (timestamp 0) glimpse location $l_t$ is at the center of the brain. After that, the glimpse moves in a direction that maximizes the reward function.

\begin{figure*}[t]
\centering
\includegraphics[width=0.25\textwidth]{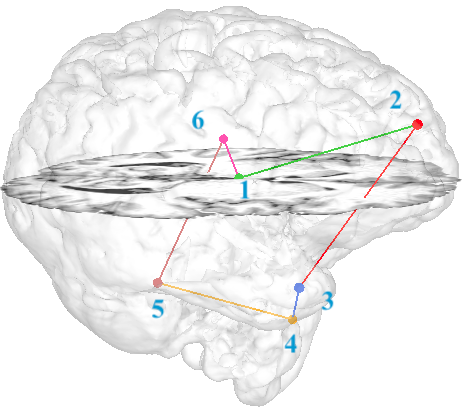}
\includegraphics[width=0.20\textwidth]{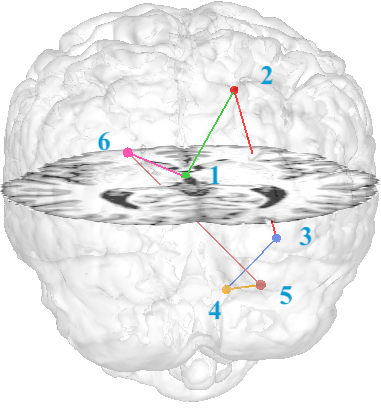}
\includegraphics[width=0.25\textwidth]{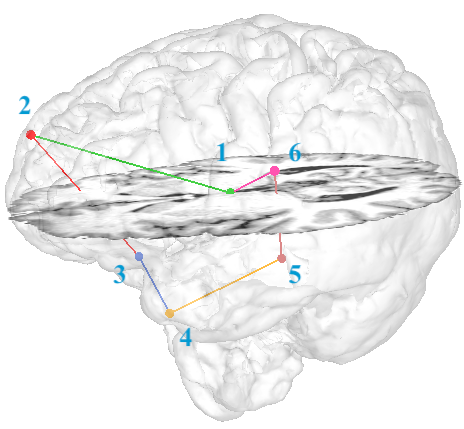}
\includegraphics[width=0.20\textwidth]{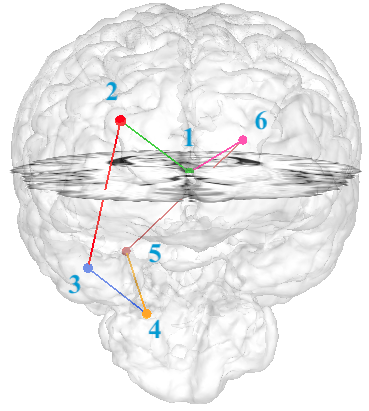} 
\caption{Trajectory taken by the glimpse algorithm. The green point is the first location, followed by the red, blue, yellow, brown, and pink points. We draw a line between them for visualization purposes.}
\label{fig:fig2}
\end{figure*}

Figure ~\ref{fig:fig2} shows the extracted glimpse locations by the 3D RVN model for a random participant with preclinical AD. The figure demonstrates how the 3D RVN agent is exploring the brain regions. The green point shows the first glimpse, which is located at the center of the brain at the first timestamp. Since we are using a total of six timestamps ($T=6$) in the current 3D RVN, there are six different areas where the model agent focuses on to make an inference. By plotting the glimpse's location, we can visualize which parts of the brain are more informative when detecting preclinical stage Alzheimer's disease.
\par 
By examining the testset brain scans, and the areas where the 3D RVN model is paying more attention (producing location $l_t$ to extract glimpse/3D image fraction), we identified that it is focusing on areas like: locus coeruleus, hippocampus, entorhinal cortex, and amygdala, parts of the brain that are important for memory and very relevant when detecting MCI \cite{memorybrain}. Another key brain region for detecting AD is the lateral ventricles \cite{ventricules}, and the 3D RVN model is focusing on that area as well. 

\subsubsection{Results for the attention transformer model.} The transformer model is trained in a different way compared to other networks. First, instead of feeding stacked 3D images to the network, we feed images in a sequence. As mentioned in the Proposed Approach section, we use a VGG16 \cite{vgg16} network as our base to extract features. We tried a pre-trained VGG16 model to see the effects of pre-trained weights, and also trained the network from scratch. As shown in Table \ref{table:prevsscratch}, the transformer model when trained from scratch achieved $4.6$\% higher F1 score.
\begin{table}[h]
\centering
\begin{tabular}[h]{lcc}
    \hline
    & Pre-trained Transformer & Our transformer \\
    \hline
    Accuracy & 88.82\% & 91.18\% \\
    F1 Score & 0.86 & 0.90 \\
    Precision & 0.91 & 0.85 \\
    Recall & 0.76 & 0.96 \\
    \hline
\end{tabular}
\caption{Comparison between pre-trained transformer and our transformer.}
\label{table:prevsscratch}
\end{table}

Due to memory limitations, we are not able to use all frames from brain MRI scans. We started by selecting 48 frames from the middle and then went up to 96 frames, which is the highest amount we can select with our current GPU configuration. We achieved the best results with 96 frames, and trained the network with same amount of images from every scan. We also trained the network for different epochs. Due to the nature of the transformer networks, we need to select higher number of epochs for an optimal training. Thus, we trained our transformer model for $200$ epochs.

As for the optimizer, we started with Adam \cite{adam}, and set the learning rate (lr) to $1e^{-4}$ and $amsgrad$ parameter to true \cite{amsgrad}. Then, we trained the network with AdamW \cite{AdamW} with $lr = 1e^{-4}$ and $amsgrad$ set as true. Finally, we trained the transformer network with SGD \cite{SGD}, and set $lr=1e^{-4}$, momentum as $9e^{-4}$, and nesterov as True \cite{pmlr-v28-sutskever13}. Among all the optimizers, we obtained better results, presented in Table \ref{tab:CMtransformer}, with SGD.

\begin{table}[h]
\resizebox{\columnwidth}{!}{%
\begin{tabular}{cc|c|c|}
\cline{3-4}
                                                 &                                                                    & \multicolumn{2}{c|}{Actual}                                                                                                   \\ \cline{2-4} 
\multicolumn{1}{c|}{}                            & Categories                                                         & \begin{tabular}[c]{@{}c@{}}Healthy /\\ other dementia\end{tabular} & \begin{tabular}[c]{@{}c@{}}Preclinical\\ AD\end{tabular} \\ \hline
\multicolumn{1}{|c|}{Predicted} & \begin{tabular}[c]{@{}c@{}}Healthy /\\ other dementia\end{tabular} & 87                                                                 & 12                                                       \\ \cline{2-4} 
\multicolumn{1}{|c|}{}                           & \begin{tabular}[c]{@{}c@{}}Preclinical\\ AD\end{tabular}           & 3                                                                  & 68                                                       \\ \hline
\end{tabular}
}
\caption{Confusion matrix for the transformer model.}
\label{tab:CMtransformer}
\end{table}

According to the evaluation results shown in Table \ref{tab:CMtransformer}, the transformer model achieves an F1 score of $0.90$, not only improving the baseline model but also outperforming the 3D RVN model. In terms of the false negative (FN) rate, the transformer model also achieves the lowest rate with $15$\%.
\par
The transformer model also achieves the best performance in terms of the earliest detection. The clinical data of patient ID 30557 is shown in Table \ref{table4}. The patient was clinically diagnosed as ``Cognitively normal’’ from day 0000 to day 3816, and first diagnosed as ``AD dementia’’ on day 4222. OASIS 3 Dataset contains two complete MRI brain scans of the patient performed on days 1448 and 2185. Transformer model infers `preclinical stage AD’ by taking the brain scan taken on day 1448 as input. This means that the algorithm is detecting `preclinical AD' 2774 days (4222 - 1448  = 2774) before the patient is clinically diagnosed.

\begin{table}[h]
\centering
\begin{tabular}[h]{lcc}
    \hline
     Day & Diagnose \\
    \hline
    0000 & Cognitively normal  \\
    0363  & Cognitively normal \\
    0749  & Cognitively normal \\
    1076  & Cognitively normal  \\
    1464  & Cognitively normal\\
    1980  & Cognitively normal\\
    2378  & Cognitively normal\\
    3093  & Cognitively normal\\
    3452  & Cognitively normal \\
    3816  & Cognitively normal \\
    4222  & Alzheimer's disease dementia \\
    4586  & Alzheimer's disease dementia \\

    \hline
\end{tabular}
\caption{Clinical data for patient 30025.}
\label{table4}
\end{table}

\begin{table}[t]
\centering
\begin{tabular}[h]{lcccc}
    \hline
    Model & Accuracy & F1 Score & Precision & Recall \\
    \hline
    Baseline & 88.34\% & 0.83 & 0.92 & 0.75  \\
    RVN & 90.65\% & 0.87 & 0.93 & 0.81  \\
    Transformer & \textbf{91.18}\% & \textbf{0.90} & \textbf{0.96} & \textbf{0.85}  \\
    \hline
\end{tabular}
\caption{Scores on test data for binary classification.}
\label{tab:OverallComp}
\end{table}

\subsubsection{Comparison between models.} In Table \ref{tab:OverallComp}, we compare the three models by measuring accuracy, precision, recall, and F1 score on our class 1 (i.e., `preclinical stage AD'). As can be seen, the transformer model outperforms others in terms of all the measured parameters. It is worth pointing out how measuring just accuracy could not always be a good metric when comparing models. In terms of accuracy, if we compare the baseline model with our transformer algorithm, the improvement is just 3.21\%. However, with F1 score e.g., the improvement is much higher (e.g., $8.4$\% higher F1 score). This is why, to have a clear picture of which model is performing better, other statistical measures, such as F1 score, precision and recall, have to be taken into consideration.

\section{Discussion}
It is important to note that a subset of individuals with biomarker evidence of preclinical AD will not progress to developed AD dementia during their lifetime \cite{rowe2010amyloid}. Hence, a fraction of the `cognitively normal' labeled/diagnosed participants in the OASIS 3 dataset have preclinical AD biomarker, even though their disease never progressed to clinicaly diagnosed dementia stage. These individuals may contribute to the false positive (misclassification to AD) rate of our evaluation results.
\par 
\begin{figure}[h]
\centering
\includegraphics[width=0.4\textwidth]{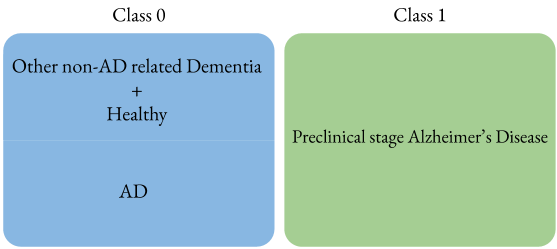} 
\caption{Overview of data labeling approach that includes AD dementia}
\label{fig:venndiagram2}
\end{figure}
In addition to using the labeling scheme shown in Fig.~\ref{fig:venndiagram1}, we performed further experiments by following the data labeling scheme shown in Fig.~\ref{fig:venndiagram2}. In this case, the `other' class included patients with diagnosed AD, and we performed classification to differentiate individuals with `preclinical AD' from others, including the patients with developed AD dementia. We compared the performance of 3D CNN and 3D RVN models. As shown in Table \ref{table:prevsscratch1}, the 3D RVN model achieves an F1 score of $0.78$ and accuracy of $84.6\%$. Even though these results are not as high compared to the previously evaluated `preclinical AD’ assessment models (using the labeling scheme in Fig.~\ref{fig:venndiagram1}), the glimpse locations identified by this RVN agent carry potential. More specifically, they highlight the brain regions responsible for differentiating  `preclinical AD' versus developed AD (stages 2 and 3). These findings will enhance our understanding on the deterioration of the brain structure with progression of AD (from preclinical to AD stages 2 and 3).

\begin{table}[h]
\centering
\begin{tabular}[h]{lcc}
    \hline
    & 3D CNN & 3D RVN \\
    \hline
    Accuracy & 83.27\% & 84.62\% \\
    F1 Score & 0.81 & 0.78 \\
    Precision & 0.94 & 0.91 \\
    Recall & 0.71 & 0.69 \\
    \hline
\end{tabular}
\caption{Binary approach to differentiate preclinical AD versus others, including developed AD.}
\label{table:prevsscratch1}
\end{table}

\section{Conclusion}
A lot of work has been done detecting MCI (stage 2), the earliest form of detectable Alzheimer's disease \cite{Jour}. The novelty and significance of our work is that  we focus on and are able to predict AD at preclinical stage (stage 1), when all the parameters, physical assessments, and clinical data state that the patient is healthy.

With that goal, we adapted two different attention-based network models, and compared their performances with a 3D CNN-based baseline that we implemented. The baseline model uses stacked MRI scans for classification. The first attention model is a recurrent attention model, which extracts glimpses from stacked images, and feeds them into recurrent attention units to get a classification result. The second model is a modified and re-purposed transformer network, which first extracts features of a sequence of images from a pre-trained network, and then feeds these features to a transformer structure to be able to classify the sequence of images. Among these three approaches, the transformer model outperformed others achieving an F1 score of $0.90$ and accuracy of 91.18\% in preclinical  (i.e., prior to MCI and dementia stage) Alzheimer disease detection. Experimental results demonstrate that, by using the MRI images effectively, it is possible to detect preclinical stage Alzheimer's disease with a very promising accuracy. 


\scriptsize
\bibliography {references}

\end{document}